\documentclass[12pt]{article}

\usepackage{latexsym,amsmath,amssymb,amsthm,amsfonts,graphicx,natbib}
\usepackage{epsfig}
\usepackage{bm,color}
\usepackage{authblk}
\newcommand*\patchAmsMathEnvironmentForLineno[1]{%
      \expandafter\let\csname old#1\expandafter\endcsname\csname #1\endcsname
      \expandafter\let\csname oldend#1\expandafter\endcsname\csname end#1\endcsname
      \renewenvironment{#1}%
         {\linenomath\csname old#1\endcsname}%
         {\csname oldend#1\endcsname\endlinenomath}}%
    \newcommand*\patchBothAmsMathEnvironmentsForLineno[1]{%
      \patchAmsMathEnvironmentForLineno{#1}%
      \patchAmsMathEnvironmentForLineno{#1*}}%
    \AtBeginDocument{%
    \patchBothAmsMathEnvironmentsForLineno{equation}%
    \patchBothAmsMathEnvironmentsForLineno{align}%
    \patchBothAmsMathEnvironmentsForLineno{flalign}%
    \patchBothAmsMathEnvironmentsForLineno{alignat}%
    \patchBothAmsMathEnvironmentsForLineno{gather}%
    \patchBothAmsMathEnvironmentsForLineno{multline}%
    }

\usepackage[mathlines,displaymath]{lineno}
\usepackage{float} 
\floatstyle{plain}
\newfloat{Algorithm}{thp}{lop}
\floatname{Algorithm}{Algorithm}
\newenvironment{wideenumerate}{\enumerate\addtolength{\itemsep}{5pt}}{\endenumerate}
\include{def}
\def\dispmuskip{\thinmuskip= 3mu plus 0mu minus 2mu \medmuskip=  4mu plus 2mu minus 2mu \thickmuskip=5mu plus 5mu minus 2mu}
\def\textmuskip{\thinmuskip= 0mu                    \medmuskip=  1mu plus 1mu minus 1mu \thickmuskip=2mu plus 3mu minus 1mu}
\def\beq{\dispmuskip\begin{equation}}    \def\eeq{\end{equation}\textmuskip}
\def\beqn{\dispmuskip\begin{displaymath}}\def\eeqn{\end{displaymath}\textmuskip}
\def\bea{\dispmuskip\begin{eqnarray}}    \def\eea{\end{eqnarray}\textmuskip}
\def\bean{\dispmuskip\begin{eqnarray*}}  \def\eean{\end{eqnarray*}\textmuskip}

\begin{document}

\title{\vspace{-0.5in}Variational Bayes with Synthetic Likelihood}
\date{\empty}

\author[1]{Victor M-H. Ong}
\author[1]{David J. Nott\thanks{Corresponding author:  standj@nus.edu.sg}}
\author[2]{Minh-Ngoc Tran}
\author[3]{S. A.  Sisson}
\author[4]{C. C. Drovandi}
\affil[1]{Department of Statistics and Applied Probability, National University of Singapore, Singapore 117546}
\affil[2]{Department of Business Analytics, The University of Sydney Business School, The University
of Sydney}
\affil[3]{School of Mathematics and Statistics, University of New South Wales, Sydney 2052 Australia}
\affil[4]{School of Mathematical Sciences, Queensland University of Technology, Brisbane 4000 Australia}

\maketitle

\vspace{-0.4in}
\begin{abstract}
Synthetic likelihood is an attractive approach to likelihood-free inference when an approximately Gaussian 
summary statistic for the data,
informative for inference about the parameters, is available.  The synthetic likelihood method derives an approximate likelihood function from 
a plug-in normal density estimate for the summary statistic, 
with plug-in mean and covariance matrix obtained by Monte Carlo simulation from the model.
In this article, we develop alternatives to Markov chain Monte Carlo implementations of Bayesian synthetic likelihoods with reduced computational overheads.
Our approach uses stochastic gradient variational inference methods 
for posterior approximation in the synthetic likelihood context, 
employing unbiased estimates of the log likelihood.  We compare the new method with a related likelihood free variational
inference technique in the literature, while at the same time improving the implementation of that approach in a number
of ways.  These new algorithms are feasible to implement in situations which are challenging for conventional approximate Bayesian computation (ABC) methods,  
in terms of the dimensionality of the parameter and summary statistic.  


\end{abstract}

\section{Introduction}\label{sec:Introduction}

Synthetic likelihood \citep{wood10} is an attractive approach to likelihood-free inference in situations where an approximately Gaussian
summary statistic for the data, informative about the parameters, is available.  As explained in \citet{price+dln16}, 
the use of synthetic likelihood mitigates to some extent
the curse of dimensionality associated with conventional approximate Bayesian computation (ABC) 
methods, and it is also convenient to apply with algorithmic parameters that are 
easy to tune.  
In this article we develop alternatives to Markov chain Monte Carlo (MCMC) implementations of Bayesian synthetic likelihoods, with reduced computational
overheads.  In particular, using unbiased estimates of the log likelihood, we implement 
stochastic gradient variational inference methods for posterior approximation that are more tolerant of noise in the likelihood estimate used.
The main contributions of this work are:  1) to improve on the variational Bayes with intractable likelihood
(VBIL) methodology of \citet{tran+nk16} by considering certain reduced variance gradient estimates, adaptive learning rates and
alternative parametrizations; 2) to modify the VBIL methodology to work with unbiased log likelihood estimates in the synthetic likelihood framework; and 3) 
to compare variational Bayes synthetic likelihood (VBSL) with pseudo-marginal MCMC synthetic likelihood implementations \citep{price+dln16}
and VBIL in a number of examples.  The new methods introduced are feasible to implement in situations which are challenging
for conventional ABC methods in terms of the dimensionality of both the parameter and summary statistic.

Suppose we have data $y$, a parameter $\theta$ of dimension $p$, a likelihood $p(y|\theta)$ which is computationally intractable, and a summary
statistic $S=S(y)$ of dimension $d\geq p$ which is assumed to be approximately Gaussian conditional on each value of $\theta$.  Inference is to be based on the observed
value $s$ of the summary statistic, which is thought to be informative about $\theta$.  
The likelihood for the summary statistic, if this statistic is assumed to be exactly Gaussian, is $\phi(s;\mu(\theta),\Sigma(\theta))$ where
$\phi(z;\mu,\Sigma)$ is the multivariate normal density with mean vector $\mu$ and covariance matrix $\Sigma$, and where $\mu(\theta)=E(S|\theta)$ and
$\Sigma(\theta)=\mbox{Cov}(S|\theta)$.  In general, however, $\mu(\theta)$ and $\Sigma(\theta)$  will be unknown.  
Synthetic likelihood \citep{wood10} replaces $\mu(\theta)$ and $\Sigma(\theta)$ by estimates obtained by simulation.  For a given $\theta$ 
we may simulate summary statistics $S_1,\dots,S_N$ under the model given $\theta$, calculate 
$$\hat{\mu}(\theta)=\frac{1}{N}\sum_{i=1}^N S_i \;\;\;\;\;\;\; \hat{\Sigma}(\theta)=\frac{1}{N-1}\sum_{i=1}^N (S_i-\hat{\mu}(\theta))(S_i-\hat{\mu}(\theta))^\top$$
and approximate $\phi(s;\mu(\theta),\Sigma(\theta))$ by
\begin{align}
 \hat{p}_N(s|\theta) & = \phi(s;\hat{\mu}(\theta),\hat{\Sigma}(\theta)).  \label{pluginsl}
\end{align}
As $N\rightarrow\infty$, $\hat{p}_N(s|\theta)$ will converge
to $\phi(s;\mu(\theta),\Sigma(\theta))$ pointwise for each value of $\theta$.  In many applications of synthetic likelihood,
users choose $N$ to be very large so that the effects of estimating $\mu(\theta)$ and $\Sigma(\theta)$ can be safely ignored.  
However, choosing $N$ large incurs a high computational cost for each synthetic likelihood evaluation.  
One way to circumvent this difficulty is to somehow emulate the synthetic likelihood, and this has been considered by a number
of authors using a variety of techniques \citep{meeds+w14,moores+dmr14,wilkinson14,gutmann+c15}.

Recently, \citet{price+dln16} considered a variation of synthetic likelihood which they call unbiased synthetic likelihood (uSL).  
In this approach (\ref{pluginsl}) is replaced by a likelihood approximation obtained from an unbiased estimate of a normal density function
due to \citet{ghurye+o69}.  
Using similar notation to \citet{ghurye+o69} let 
$$c(k,\nu)=\frac{(2\pi)^{-k\nu/2}\pi^{-k(k-1)/4}}{\prod_{i=1}^k \Gamma\left(\frac{1}{2}(\nu-i+1)\right)},$$
and for a square matrix $A$ write $\psi(A)=|A|$ if $A>0$ and $0$ otherwise, where $|A|$ is the determinant of $A$ and $A>0$ means that $A$ is positive definite.
Then in uSL (\ref{pluginsl}) is replaced by
\begin{align}
  \hat{p}_N^U(s|\theta) & = (2\pi)^{-\frac{d}{2}}\frac{c(d,N-2)}{c(d,N-1)(1-1/N)^{d/2}}\left|S_\theta\right|^{-\frac{N-d-2}{2}}\psi\left(S_\theta-\frac{(s-\hat{\mu}(\theta))(s-\hat{\mu}(\theta))^\top}{(1-1/N)}\right)^{\frac{N-d-3}{2}}, \label{usl}
\end{align}
where $S_\theta=(N-1)\hat{\Sigma}(\theta)$.  
The results of \citet{ghurye+o69} imply that $E(\hat{p}_N^U(s|\theta))=\phi(s;\mu(\theta),\Sigma(\theta))$ 
if the summary statistic is Gaussian, provided that $N>d+3$.   
This unbiasedness property means that if (\ref{usl}) is used in a pseudo-marginal MCMC algorithm 
\citep{beaumont03,andrieu+r09} and if $S$ is actually normally distributed, then 
the Markov chain converges
to the exact posterior regardless of the value of $N$.  However, even though the distribution targeted by such a pseudo-marginal algorithm
does not depend on $N$, the mixing of the algorithm can be very poor unless $N$ is chosen large enough to control the variance of the likelihood
estimate.   \citet{doucet+pdk15} 
suggest fixing the variance of the log likelihood estimate to be around 1 for pseudo-marginal Metropolis-Hastings algorithms, to achieve an optimal trade off between computational cost and precision.

An alternative approach to MCMC
methods for Bayesian computation is variational approximation (see, for example, \citet{bishop06} and \citet{ormerod+w10}).  Although variational approximation
is an approximate
inference method, it can often be implemented with an order of magnitude less computational effort than the corresponding ``exact" algorithms such as MCMC.  
Recently, \citet{tran+nk16} considered the use of stochastic gradient variational inference when the likelihood is computationally intractable, and 
only an unbiased estimate of the likelihood is available.  This includes situations 
where conventional ABC methods \citep{marin+prr12,blum+nps13} are usually applied.  
In standard ABC, a nonparametric approximation to the likelihood is used.
With $K_\epsilon(\cdot,\cdot)$ a kernel function in which $\epsilon>0$ is a bandwidth parameter, ABC considers the likelihood approximation
\begin{align}
 \tilde{p}(s|\theta) & =\int K_\epsilon(s,S(y'))p(y'|\theta)dy'  \label{abclike}
\end{align}
which is estimated unbiasedly by
\begin{align}\label{ABC_like}
 \hat{p}(s|\theta) & = \frac{1}{N}\sum_{i=1}^N K_\epsilon(s,S(y_i'))
\end{align}
where $y_1',\dots,y_N'$ are iid draws from $p(y|\theta)$.  

In principle, we can use the estimate (\ref{usl}) to give a synthetic likelihood version of the VBIL method of
\citet{tran+nk16} -- this is discussed further in Section \ref{sec:vbil}.  This may be beneficial compared to unbiased estimation of (\ref{abclike}), 
since the parametric assumptions made in the synthetic likelihood mean that 
the synthetic likelihood can be estimated more precisely for a given number of model simulations, $N$,  than  the corresponding ABC likelihood.  
However, for implementing stochastic gradient variational Bayes (VB) methods, it is much more convenient
to work with unbiased estimates of the log likelihood function (see Section \ref{sec:uslle}).  Unbiased estimation of the log likelihood corresponding to (\ref{abclike}) cannot be achieved directly.   
Furthermore, the VBIL method using an unbiased likelihood estimate is not easy to apply in some ABC problems, 
as the user needs  to tune the variance of the log-likelihood
estimator to be constant across the parameter space -- see Section \ref{sec:vbil} for further details.  
However, stochastic gradient VB methods, which use unbiased estimates of a log likelihood, have no such requirement.  
Unbiased estimators of the log of a normal density function are available
from the pattern recognition literature (\citet{ripley96}, p. 56). Hence, assuming that the summary statistic is Gaussian, unbiased estimates of the log likelihood are available 
in the synthetic likelihood context.
This makes the implementation of stochastic gradient VB methods very easy.  

The next section reviews stochastic gradient VB methods, and Section \ref{sec:vbil} explains the VBIL method
of \citet{tran+nk16}.  Our VBSL algorithm is described in Section 4, as well as some refinements of the basic stochastic gradient
optimization approach that apply both to VBIL and VBSL.  Section 5 
compares VBSL with VBIL and pseudo-marginal synthetic likelihood approaches in some challenging examples. We conclude with a discussion.

\section{Stochastic gradient variational Bayes}\label{sec:sgvb}

Consider a Bayesian inference problem with data $y$, a $p$-dimensional parameter $\theta$, prior  distribution $p(\theta)$
and likelihood function $p(y|\theta)$, so that the posterior density is $p(\theta|y)\propto p(\theta)p(y|\theta)$.  
In variational inference the posterior density is approximated by a density within some tractable family.  Here we consider a parametric
family with typical element $q_\lambda(\theta)$, where $\lambda$ is a variational parameter to be chosen.  
The Kullback-Leibler divergence from $q_\lambda(\theta)$ to $p(\theta|y)$ is given by
\begin{align}
  KL(\lambda) 
  = KL(q_\lambda(\theta) || p(\theta|y)) 
 = \int \log \frac{q_\lambda(\theta)}{p(\theta|y)} q_\lambda(\theta) d\theta.  \label{kld}
\end{align}
Denote the marginal likelihood by $p(y)=\int p(\theta)p(y|\theta)d\theta$.  
Minimizing $KL(\lambda)$ with respect to $\lambda$ is equivalent to maximizing
\begin{align*}
  {\cal L}(\lambda) & = \int \log \frac{p(\theta)p(y|\theta)}{q_\lambda(\theta)} q_\lambda(\theta) d\theta,
\end{align*}
 and it can be shown that ${\cal L}(\lambda)$ is a lower bound on the log marginal likelihood $\log p(y)$.
For introductory discussion of VB methods see e.g.  \citet{bishop06} and \citet{ormerod+w10}.
In non-conjugate settings  ${\cal L}(\lambda)$ may not be directly computable.  In this setting, stochastic gradient methods \citep{robbins+m51,bottou10}
have been developed which can optimize ${\cal L}(\lambda)$ effectively even when it can't be calculated analytically, provided simulation from $q_\lambda(\theta)$ is
possible \citep{ji+sw10,nott+tvk12,paisley+bj12,salimans+k13,kingma+w13,hoffman+bwp13,rezende+mw14,titsias+l14,titsias+l15}.

The most general approaches to using stochastic gradient methods in VB have been based on the ``log derivative trick".  
Observe that $\nabla_\lambda q_\lambda(\theta)=q_\lambda(\theta)\nabla_\lambda \log q_\lambda(\theta)$, and that $E(\nabla_\lambda \log q_\lambda(\theta))=0$ (where the expectation is with respect to $q_\lambda(\theta)$).  This last identity follows from differentiating both sides of the equation $\int q_\lambda(\theta)d\theta=1$ with respect to $\lambda$.  Writing $h(\theta)= p(\theta)p(y|\theta)$, then
\begin{align}
 \nabla_\lambda {\cal L}(\lambda) & = \nabla_\lambda \int \left\{ \log h(\theta)-\log q_\lambda(\theta) \right\} q_\lambda(\theta) d\theta  \nonumber \\
 & = \int \log h(\theta)\nabla_\lambda \log q_\lambda(\theta) q_\lambda(\theta)d\theta - \int \log q_\lambda(\theta)\nabla_\lambda \log q_\lambda(\theta) q_\lambda(\theta)d\theta \nonumber \\
 & = \int \nabla_\lambda \log q_\lambda(\theta) \left\{ \log h(\theta)-\log q_\lambda(\theta)\right\} q_\lambda(\theta) d\theta.  \label{lbgradient}
\end{align}
The last expression is an expectation with respect to $q_\lambda(\theta)$, which is easily estimated unbiasedly if we can simulate from $q_\lambda(\theta)$.  This then permits implementation
of a stochastic gradient algorithm for optimizing ${\cal L}(\lambda)$.  
In the original lower bound expression, some terms (e.g. $E(\log q_\lambda(\theta))$) can sometimes be calculated analytically, in which case
the estimate (\ref{lbgradient}) can be modified appropriately, although this may not always be beneficial  \citep{salimans+k13}.  
It is well known that gradient estimates obtained by the log derivative trick are highly variable, and a variety of additional methods for variance reduction 
have also been considered in the above references.  \citet{titsias+l15} recently considered an interesting approach that can be implemented in a model independent
fashion.  

For large datasets it is convenient to replace the log likelihood term in $\log h(\theta)$ by an unbiased estimate -- this still results in an unbiased estimate
of the gradient of ${\cal L}(\lambda)$.  Such estimates of the log-likelihood are usually obtained by subsampling. Variational schemes that use
both subsampling and sampling from the variational posterior to generate gradient estimates have been termed ``doubly stochastic" by \citet{titsias+l14} (see also \citet{kingma+w13} and \citet{salimans+k13} for similar approaches).  The variational Bayes with intractable log likelihood (VBILL) methodology of \citet{gunawan+tk16} considers 
unbiased estimation of log likelihoods within stochastic gradient variational inference using difference estimators for variance reduction.

\section{Variational Bayes with intractable likelihood (VBIL)}\label{sec:vbil}

We now describe the VBIL method of \citet{tran+nk16} since we build on this approach in Section 4.  
VBIL is the first attempt to apply stochastic gradient variational inference methods to a class of problems that includes
 likelihood-free inference, and uses black box variational inference methods \citep{ranganath+gm14}.  
However, a related expectation propagation approach to likelihood free inference has been considered previously by 
\citet{barthelme+c14}.  More recently \citet{moreno+amrw16} have considered an automatic variational ABC approach 
based on stochastic gradient VB 
with attractive methods for gradient estimation, which apply when the forward simulation model can be written 
as a differentiable function of both model parameters and random variables, and when the model code is written in an automatic differentiation
environment.  

The VBIL approach works with an unbiased estimate of the likelihood which we denote by
$\hat{p}_N(y|\theta)$.  Here $N$ is an algorithmic parameter controlling the accuracy of the approximation, such as the number of Monte Carlo samples used.  
Following \citet{pitt+sgk12} and \citet{tran+nk16} we refer to $N$ as the number of particles.  Write $z=\log \hat{p}_N(y|\theta)-\log p(y|\theta)$, and 
$g_N(z|\theta)$ for the distribution of $z$ given $\theta$.  Since $\hat{p}_N(y|\theta)$ is unbiased, we must have
\begin{align}
 \int \exp(z)g_N(z|\theta)d\theta & =1.  \label{unbiasedness}
\end{align}
\citet{tran+nk16} consider implementing VB in the augmented space $(\theta,z)$, inspired by similar ideas in the literature on pseudo-marginal MCMC
algorithms \citep{beaumont03,andrieu+r09}, and
in particular, consider the target
distribution $p_N(\theta,z)=p(\theta|y)\exp(z)g_N(z|\theta)$.  Using (\ref{unbiasedness}), we see that 
the $\theta$ marginal of $p_N(\theta,z)$ is the posterior distribution of interest, 
$p(\theta|y)$.  Consider a family of approximating distributions of the form
$$q_\lambda(\theta,z)=q_\lambda(\theta)g_N(z|\theta)$$
where $\lambda$ is a variational parameter to be chosen.  The $\theta$ marginal of $q_\lambda(\theta,z)$ is $q_\lambda(\theta)$.  
Performing the VB optimization in the augmented space, by choosing $\lambda$ to minimize $KL(q_\lambda(\theta,z) || p_N(\theta,z))$,  then  
the gradient of the objective function can be shown to be
\begin{align}
 & E(\nabla_\lambda \log q_\lambda(\theta)(\log(p(\theta)\hat{p}_N(y|\theta))-\log q_\lambda(\theta))) \label{vbilgradient}
\end{align}
where the expectation is with respect to $q_\lambda(\theta,z)$. The expression in (\ref{vbilgradient}) is easily obtained from
(\ref{lbgradient}), and 
 is easily approximated by simulation, since all that is required
is simulation of $\theta$ from $q_\lambda(\theta)$ and calculation of the likelihood estimate $\hat{p}_N(y|\theta)$.
Knowledge of $z$, which depends on the unknown $p(y|\theta)$, is not required.

Minimization of $KL(q_\lambda(\theta,z) || p_N(\theta,z))$ is not the same in general as minimization of $KL(\lambda)$ given by (\ref{kld}).  However, 
\citet{tran+nk16} show that if a) there is a function $\gamma^2(\theta)>0$ such that $E(z|\theta)=-\gamma^2(\theta)/(2N)$ and 
$\mbox{Var}(z|\theta)=\gamma^2(\theta)/N$, and b) for a given $\sigma^2>0$, $N$ can be chosen as a function of $\theta$ and $\sigma^2$ so that
$\mbox{Var}(z|\theta)\equiv \sigma^2$, then the minimizers of $KL(q_\lambda(\theta,z) || p_N(\theta,z))$ and $KL(\lambda)$ correspond.
The lower bound in the augmented space is
\begin{align*}
  {\cal L}_a(\lambda)=\int \log \frac{p(\theta)p(y|\theta)\exp(z)g_N(z|\theta)}{q_\lambda(\theta)g_N(z|\theta)} q_\lambda(\theta,z)=& 
{\cal L}(\lambda)+\int z g_N(z|\theta)q_\lambda(\theta)d\theta
\end{align*}
which is ${\cal L}(\lambda)$ plus a constant which is independent of $\lambda$ if $N$ has been tuned so that $E(z|\theta)$ does not depend on $\theta$.  
If the log likelihood estimator is asymptotically normal, so that $z$ is normal, this implies that asymptotically $z|\theta\sim N(E(z|\theta),-2E(z|\theta))$ by the unbiasedness
condition.  
Hence, tuning $E(z|\theta)$ to not depend on $\theta$ is equivalent to tuning the variance of the log-likelihood estimator to not depend on $\theta$
in this case. The resulting lower bound in the augmented space is
\begin{align}
  {\cal L}_a(\lambda) & = {\cal L}(\lambda)-\frac{\tau^2}{2}  \label{auglowerbound}
\end{align}
where $\tau^2$ is the targeted variance for the log-likelihood estimator.  
\citet{tran+nk16} show that this approach is more tolerant of noise in the likelihood
estimate than pseudo-marginal MCMC algorithms which use similar unbiased estimates of the likelihood.  

The VBIL method of \citet{tran+nk16} is useful in a number of settings, such as state space models and random effects models, 
where it is convenient to obtain unbiased estimates of the likelihood.
It is also useful for ABC since it is trivial to estimate (\ref{abclike}) unbiasedly. 
Crucial to the VBIL method is the use of variance reduction methods in the gradient estimates in the stochastic
gradient procedure.  In this article, we consider only
multivariate normal approximations to the posterior;  exploiting the fact that such approximations are in the exponential family allows
the use of natural gradient methods \citep{amari98} as described in \citet{tran+nk16}.  
Using these ideas as well as the control variates approach to variance reduction described in \citet{tran+nk16} results in 
Algorithm 1.  Further justifications for the details of the algorithm are given in Section 3 of \citet{newtran+nk16}.
In Algorithm 1, $\lambda$ denotes the natural parameters in the normal variational posterior distribution $q_\lambda(\theta)$ and
$I_F(\lambda)=\mbox{Cov}(\nabla_\lambda \log q_\lambda(\theta))$.  Details of the parametrization and form of $I_F(\lambda)$ 
are given in Appendix A.  In Algorithm \ref{Alg1} we also write $n$ for a sample size parameter that scales the lower bound, and 
$S$ is the number of samples used in the gradient estimate.  
Finally, $\rho_t$, $t\geq 0$, is a learning rate sequence satisfying the Robbins-Monro conditions  $\sum_t \rho_t=\infty$,
$\sum_t \rho_t^2<\infty$ \citep{robbins+m51}.   

We note that there are two differences between Algorithm 1 based on \cite{newtran+nk16}, and the earlier approach described in \citet{tran+nk16}.  
Firstly, it is suggested in \citet{newtran+nk16} that the values $\theta^{(s)}$, $s=1,\dots,S$ in step 1 can be generated using
randomized quasi Monte Carlo, and this can be helpful for reducing the variance of the gradient estimates in some problems.  
Secondly, 
Algorithm 1 follows 
\citet{newtran+nk16} in 
estimating all parts of the lower bound expression using Monte Carlo with the same $\theta$ samples
to reduce variance of gradient estimates, rather than
calculating certain parts of the lower bound analytically (see \citet{newtran+nk16} for further discussion).

\begin{Algorithm*}
\centering
\parbox{0.8\textwidth}{
\hrule
\vspace{1mm}
Initialize $\lambda^{(0)}=(\lambda_1^{(0)},\lambda_2^{(0)})$, $t=0$, 
$\lambda^{(1)}=\lambda^{(0)}$.  $N$ is the number of particles, $S$ the number
of $\theta$ samples used in the gradient estimates.\\
\begin{wideenumerate}
\item 
\begin{wideenumerate}
\item Generate $(\theta^{(t)},z_s^{(t)})\sim q_{\lambda^{(t)},N}(\theta,z)$, $s=1,\dots, S$.  Note that the $z_s^{(t)}$ can be generated only implicitly through computation of estimates
$\hat{p}_N^S(y|\theta^{(t)})$, $s=1,\dots,S$.  
\item Set 
$$c^{(t)}=\frac{\mbox{Cov}(\hat{h}(\theta,z)\nabla_\lambda \log q_\lambda(\theta),\nabla_\lambda \log q_\lambda(\theta))}{\mbox{Var}(\nabla_\lambda \log q_\lambda(\theta))}$$
where $\mbox{Cov}(\cdot)$ and $\mbox{Var}(\cdot)$ are sample estimates of covariance and variance based on the samples $(\theta_s^{(t)},z_s^{(t)})$, $s=1,\dots,S$, 
and $\hat{h}(\theta,z)=\log p(\theta)\hat{p}_N(y|\theta)$.    
\item $t=t+1$.
\end{wideenumerate}
\item Repeat
\begin{wideenumerate}
\item Generate $(\theta_s^{(t)},z_s^{(t)})\sim q_{\lambda^{(t)},N}(\theta,z)$, $s=1,\ldots, S$.
\item $\hat{H}^{(t)}=\frac{1}{S}\sum_{s=1}^S (\hat{h}(\theta_s^{(t)},z_s^{(t)})-\log q_\lambda(\theta_s^{(t)})-c^{(t-1)}) \nabla_\lambda \log q_\lambda(\theta_s^{(t)})$.
\item Estimate $c^{(t)}$ as in step 1 (b).  
\item $\tilde{\lambda}^{(t+1)}=\lambda^{(t)}+\rho_t I_F(\lambda^{(t)})^{-1}\hat{H}^{(t)}$
\item If $\Sigma(\tilde{\lambda}^{(t+1)})$ is not positive definite $\lambda^{(t+1)}=\lambda^{(t)}$ else $\lambda^{(t+1)}=\tilde{\lambda}^{(t+1)}$.  
\item Set $LB^{(t)}=\left\{\frac{1}{S} \sum_{s=1}^S \hat{h}(\theta_s^{(t)},z_s^{(t)}) -\log q_{\lambda^{(t)}}(\theta_s^{(t)}) \right\}$.
\item $t=t+1$
\end{wideenumerate}
\end{wideenumerate}
until some stopping rule is satisfied.
\vspace{1mm}
\hrule}
\caption{\small VBIL algorithm with Gaussian variational posterior distribution. Further details of the parametrisation of the variational distribution and computation of $I_F(\lambda)$ are provided in the Appendix.}\label{Alg1}
\end{Algorithm*} 
In Algorithm 1, $N$ is treated as fixed.  However, we would like $N$ to be chosen adaptively so that the variance of
the log likelihood estimator is approximately constant with $\theta$ (or at least approximately constant over the high posterior probability region).  
Hence,  in practice we adapt $N$ by first setting some minimum value $N'$  for the number of simulations in the likelihood estimation.  
Then, if some target value for the log likelihood variance is exceeded based on an empirical estimate, 
an additional number of particles (50, say) is repeatedly simulated, until the target accuracy is achieved.
This adaptive procedure does not bias the likelihood estimate obtained.


\section {Variational Bayes synthetic likelihood (VBSL)}\label{VBSLmain}

We now consider some extensions of Algorithm \ref{Alg1} --  in particular, we 
incorporate the use of the synthetic likelihood, resulting in the VBSL algorithm.  Additionally, we 
develop an adaptive method for determining the algorithm learning rates, and 
reparametrizations that may be helpful in cases where ensuring the positive definiteness of the variational posterior covariance matrix
is difficult.

\subsection{Unbiased synthetic log likelihood estimation}  
\label{sec:uslle}

Following \citet[p. 56]{ripley96},  when the summary statistics are normally distributed, an 
unbiased estimate of the log of a normal density $\log \phi(s;\mu(\theta),\Sigma(\theta))$ based on a random sample of size $N$ from it leading
to sample mean and covariance matrix $\hat{\mu}(\theta)$ and $\hat{\Sigma}(\theta)$ respectively is \begin{align}
 \hat{l}_N^U(s|\theta) & =-\frac{d}{2}\log 2\pi-\frac{1}{2}\left\{\log |\hat{\Sigma}(\theta)|+ d\log\left(\frac{N-1}{2}\right)-\sum_{i=1}^d \psi\left(\frac{N-i}{2}\right)\right\} \nonumber \\
  & \hspace{1.25in}-\frac{1}{2}\left\{\frac{N-d-2}{N-1}(s-\hat{\mu}(\theta))^T\hat{\Sigma}(\theta)^{-1}(s-\hat{\mu}(\theta)) - \frac{d}{N} \right\} \label{ulsl}
\end{align}
provided that $N>d+2$, where $\psi(\cdot)$ denotes the digamma function.  Hence although unbiased estimation of the logarithm of (\ref{abclike}) for the nonparametric ABC likelihood
approximation cannot be achieved directly, in the context of synthetic likelihood, where the summary statistic is assumed to follow a Gaussian distribution,
it is straightforward to use (\ref{ulsl}) as an unbiased estimate of the log likelihood.  
To implement a stochastic gradient VB algorithm for approximation of the posterior, the only change required in Algorithm 1 is to
replace $\log \hat{p}_N(y|\theta)$ wherever it appears by the expression (\ref{ulsl}) above.  

However, note that the previous requirements for minimisation of $KL(q_\lambda(\theta,z) || p_N(\theta,z))$ to correspond to minimisation of $KL(\lambda)$
in VBIL can now be dropped -- it is no longer necessary to tune $N$ as a function of $\theta$ so that the variance of the log likelihood estimator
is approximately constant.  In addition, the parametric assumptions used in the synthetic likelihood enable us 
to both reduce the variance of the log likelihood estimator for a given number of simulations, and also that of the  stochastic gradients in Algorithm 1 and our refinements.  

In many situations the assumptions made in the synthetic likelihood are reasonable -- the statistics can often be chosen, perhaps after transformation, so that they satisfy some central
limit theorem
\citep{wood10}.  
\citet{price+dln16} find that the Bayesian synthetic likelihood posterior generally seems to be not very sensitive to violations of the Gaussian assumption.
The synthetic likelihood approach may be particularly helpful for large datasets where the forward model simulations are expensive.  For large datasets
the normal variational posterior approximation will often be very reasonable, as well as the normal distributional assumption of the summary statistics.
The VBSL approach can work very efficiently in this situation without much loss of accuracy.  

Perhaps the most important advantage of the VBSL algorithm, however, is that it's tuning parameters are much easier to set than for VBIL.  In particular, for VBIL the ABC tolerance $\epsilon$ must be chosen beforehand, and in general the accuracy of the approximation as well as the variance of the gradient estimates within
the algorithm are very sensitive to this choice. Practically, as a result, multiple implementations of VBIL with different $\epsilon$ values will be required 
to establish a reasonable computation time and accuracy trade off.  The analogous parameter in the VBSL algorithm is $N$, the number of Monte Carlo samples used
in the empirical estimation of the mean and covariance matrix of the summary statistics.  If the summary statistic is exactly Gaussian distributed, the solution to the variational optimization
problem does not depend on $N$, and in practice, if the distribution is close to Gaussian there is very little sensitivity to this choice.  

\subsection{Adaptive learning rate}

A second refinement of Algorithm 1 applicable to both VBSL and VBIL is to use an adaptive learning rate.  In \citet{tran+nk16} the learning rate $\rho_t$ is chosen
to be some sequence satisfying the Robbins-Monro conditions $\sum_t \rho_t=\infty$, $\sum_t \rho_t^2<\infty$ where 
the sequence has a specified form with parameters that need to be manually tuned.  However, suitable adaptive choices
of the step sizes can avoid manual tuning, improve  convergence and make algorithm stability and performance less
sensitive to starting values.  We propose an adaptive learning rate choice based on previous work by 
\citet{ranganath+wbx13} in the context of stochastic variational inference (SVI) \citep{hoffman+bwp13}.  Similar to Algorithm 1, SVI is a stochastic natural gradient
ascent algorithm, but one where the stochasticity of the gradient estimates derive from subsampling.  The arguments provided by \citet{ranganath+wbx13}
justifying their adaptive learning rate carry over to the current setting, where the stochasticity in the estimate of the natural gradient comes
from sampling the variational distribution and from estimation of the log likelihood itself.  

Let $\hat{n}_t$ be the natural gradient estimate for the lower bound at time $t$, $\hat{n}_t=I_F(\lambda^{(t)})^{-1}\hat{H}^{(t)}$.  
A running average of the values of $\hat{n}_t$ and $\hat{n}_t^\top\hat{n}_t$ can be maintained as
\begin{align*}
 \bar{n}_t= & (1-\alpha_t)\bar{n}_{t-1}+\alpha_t \hat{n}_t \\
 \bar{c}_t= & (1-\alpha_t)\bar{c}_{t-1}+\alpha_t \hat{n}_t^T\hat{n}_t,
\end{align*}
where $\alpha_t$ is a discounting factor. 
The learning rate $\rho_t$ is then given by
\begin{align*}
 \rho_t = & \frac{\bar{n}_t^\top \bar{n}_t}{\bar{c}_t},
\end{align*}
with $\alpha_t$ also adapted as 
\begin{align*}
 \alpha_{t+1}^{-1}= & \alpha_t^{-1}(1-\rho_t)+1.
\end{align*}
The initial values $\bar{n}_0$ and $\bar{c}_0$ are chosen based on computation of $K$ independent gradient estimates at the starting value for the variational
parameters, and $\alpha_0$ is initialised as $1/K$.  
Intuition behind the choice of $\rho_t$ is that $\bar{n}_t^\top\bar{n}_t$ represents the ``signal" in the noisy gradient estimates, whereas $\bar{c}_t$ 
represents the extent of the total variation, including both signal and noise.  So large steps will be taken when the magnitude of the gradient is large compared to the noise,
whereas if the noise dominates the signal small steps are chosen.  The adaptation of the discounting factors $\alpha_t$ is implemented in such a way that more weight is given
to the current iteration following a big step.  
The rationale for the approach is based on minimising some loss function, which measures how well one step of the approach mimics the approach with noise free gradient (see \citet{ranganath+wbx13} for further discussion).  
However, we find that in some of our applications, using the proposed adaptive learning rate may still lead to instability at early iterations. We find it helpful
to set a maximum step size in the early iterations, which in our examples we choose as $\rho_t\leq \sqrt{d/\bar{c}_t}$.  

\subsection{Cholesky parametrisation of the covariance matrix}\label{Repara_update}

Our final modification of Algorithm 1 is to parametrise  the normal variational distribution
in terms of the Cholesky factor of the precision matrix.  
Implementing natural gradient steps can still be performed conveniently for this parametrisation.  
In the natural parametrisation of the normal distribution used in Algorithm 1, 
it is possible for an update to result in a parameter value $\lambda$ for which $\Sigma$ is not positive definite. In Algorithm 1 such updates are rejected, however for high-dimensional problems and with poor choices of starting values or noisy gradients, such rejection steps may occur
frequently resulting in slow convergence.  
Reparametrisation in terms of the Cholesky factor avoids this. 

In describing the implementation of the Cholesky parameterisation we require some notation, similar to that found in 
\citet{magnus+n99} and \citet{wand14}.  
For a $d\times d$ matrix $A$, write $\mbox{vec}(A)$ for the vector of
length $d^2$ obtained by stacking the columns one underneath another moving form left to right. When $A$ is symmetric, write $\mbox{vech}(A)$ for the vector with $d(d+1)/2$ elements obtained by stacking the lower triangular elements of $A$. 

We parametrise
the normal variational posterior distribution in terms of the mean $\mu$ and the (lower triangular) Cholesky factor $C$ of $\Sigma^{-1}$ so that
$\Sigma^{-1}=C C^\top$.  We do not enforce the constraint that the diagonal elements of $C$ be positive as such non-uniqueness is not
a concern in the present context.  Our variational parameters are now
\begin{align}
 \lambda= & \left[\begin{array}{cc} \mu \\ \mbox{vech}(C) \end{array} \right]. \label{newpar}
\end{align} 

We then have
\begin{align*}
 \log q_\lambda(\theta)= & -\frac{d}{2}\log2\pi-\log |C|-\frac{1}{2}(\theta-\mu)^T C C^\top (\theta-\mu)
\end{align*}
and upon differentiation with respect to $\mu$ and $\mbox{vech}(C)$ 
\begin{align*}
  \nabla_\lambda \log q_\lambda(\theta) = & \left[\begin{array}{cc} CC^\top(\theta-\mu) \\ \mbox{vech}(\mbox{diag}(1/C)-(\theta-\mu)(\theta-\mu)^\top C) \end{array}\right],
\end{align*}
where $\mbox{diag}(1/C)$ denotes the diagonal matrix with the same dimensions as $C$ with $i$th diagonal entry $1/C_{ii}$.  
This expression for $\nabla_\lambda \log q_\lambda(\theta)$ allows us to construct an unbiased gradient estimate from (\ref{lbgradient}).  However, 
Algorithm 1 uses the natural gradient, and we would like to construct a natural gradient algorithm in the new parametrisation.  
To do this we need $I_F(\lambda)=\mbox{Cov}_\lambda(\nabla_\lambda \log q_\lambda(\theta))$.  
Writing $I_F(\lambda)$ in block form, corresponding to the partition in (\ref{newpar}), then
\begin{align*}
 I_F(\lambda) = & \left[\begin{array}{cc} I_{11}(\lambda) & I_{21}(\lambda)^\top \\ I_{21}(\lambda) & I_{22}(\lambda) \end{array}\right].
\end{align*}
Write $L_d$ for the elimination matrix of order $d$ \citep{magnus+n99} which for a (not necessarily symmetric) $d\times d$ matrix $A$, transforms $\mbox{vec}(A)$ into 
$\mbox{vech}(A)$, and write $\otimes$ for the Kronecker product. We also denote by $D_d$ the duplication matrix of order $d$, which is the unique
$d^2\times d(d+1)/2$ matrix of zeros and ones such that
$$D_d\mbox{vech}(A)=\mbox{vec}(A)$$ for symmetric $d\times d$ matrices $A$, 
and its Moore-Penrose inverse is written as $D_d^+=(D_d^\top D_d)^{-1}D_d^\top$.   Then, we get
\begin{align*}
 I_{22}(\lambda) & =\mbox{Cov}(\mbox{vech}((\theta-\mu)(\theta-\mu)^\top C)) \\
 & = \mbox{Cov}(L_d\mbox{vec}((\theta-\mu)(\theta-\mu)^\top C)) \\
 & = \mbox{Cov}(L_d (C^T\otimes I)\mbox{vec}((\theta-\mu)(\theta-\mu)^\top )) \\
 & = L_d (C^\top\otimes I)\mbox{Cov}(\mbox{vec}((\theta-\mu)(\theta-\mu)^T)(C\otimes I){L_d}^\top \\
 & = L_d (C^\top\otimes I)D_d \mbox{Cov}(\mbox{vech}((\theta-\mu)(\theta-\mu)^T))D_d^T (C\otimes I) {L_d}^\top \\
 & = 2 L_d (C^\top\otimes I) D_d D_d^+ (\Sigma \otimes \Sigma) {D_d^+}^T D_d^T (C\otimes I) {L_d}^\top 
\end{align*}
where in the final line we have used the expression for $\mbox{Cov}(\mbox{vech}(xx^\top))$ for normal $x$ derived in the proof
of Theorem 1 c) of \citet{wand14}.  Finally
$$I_{11}(\lambda)=\mbox{Cov}(CC^\top(\theta-\mu))=CC^\top \Sigma CC^\top=CC^\top=\Sigma^{-1},$$
and
\begin{align*}
 I_{21}(\lambda) & = -\mbox{Cov}(\mbox{vech}((\theta-\mu)(\theta-\mu)^\top C),CC^\top(\theta-\mu)) \\
 & = L_d \mbox{Cov}(\mbox{vec}((\theta-\mu)(\theta-\mu)^T C), \theta-\mu) CC^\top \\
 & = -L_d \mbox{Cov}((C^\top\otimes I)\mbox{vec}((\theta-\mu)(\theta-\mu)^T),\theta-\mu)CC^\top \\
 & = - L_d(C^\top\otimes I) \mbox{Cov}(\mbox{vec}((\theta-\mu)(\theta-\mu)^T),\theta-\mu)CC^\top \\
 & = 0,
\end{align*}
where in the last line we have used the fact that odd order central moments of the multivariate normal distribution
are zero.  That is,  we can compute $I_F(\lambda)$ in the new parametrisation, allowing for a natural gradient implementation.
In our application in Section \ref{example:gandk} we directly compare the natural gradient approach with the use of the ordinary gradient
in the Cholesky parametrisation, with a
 per parameter adaptive learning rate determined according to the ADADELTA approach of \citet{zeiler12}. 
%
%

\section{Applications}

We investigate the performance of the VBSL approach using four different models. In the first experiment, we consider a toy example using data generated  from a Gaussian distribution.  This example permits direct comparison with the VBIL method, since the calculations can be performed analytically, and the effects of 
the finite ABC tolerance $\epsilon$ can be separated from the inaccuracy of the variational approximation itself in the VBIL algorithm.  
In the next two examples, we investigate $\alpha$-stable and multivariate $g$-and-$k$ models, which do not have closed form expressions for the density.  
The $\alpha$-stable analysis is used to demonstrate the importance of adaptive learning rates, and the $g$-and-$k$ analysis is used to compare our adaptive natural gradient optimisation scheme with a method based on the
ordinary gradient and an adaptive per parameter learning rate (the ADADELTA method of \citet{zeiler12}).      Since the multivariate $g$-and-$k$ model possesses
a fairly high-dimensional parameter, it gives some insight into how to implement the VBSL methodology in an efficient and stable way in this setting.  
Finally, our last example considers the case of a very high-dimensional summary statistic, using a real problem from cell biology. 

\subsection{Toy Example - Normal Location Model}

We consider data, $y_1,...,y_n$, from a Gaussian distribution with unknown mean $\theta$ and unit variance.  
We assume that the observed data is $y=(0,\dots,0)$ and adopt
a standard normal distribution $N(0,1)$ for the prior on $\theta$ so that the posterior distribution is $\theta | y \sim N( n/(1+n) \bar{y}, 1/(1+n))$ where $\bar{y}$ denotes the sample mean. 
We ignore the fact  that $\bar{y}$ is a sufficient statistic and take the entire data set $y$ as the summary statistic.  
This allows us to explore the effect of increasing dimension of the summary statistic on the likelihood free methods.  
For the VBIL approach, we use the ABC likelihood \eqref{ABC_like} with a Gaussian kernel defined as 
\begin{align}\label{ABC_normal}
K_{\epsilon} (s,s') = (2\pi \epsilon)^{-d/2} \exp \left\{ -\frac{1}{2\epsilon } [ (s -s')^\top (s -s')   ] \right\}.
\end{align}
 With this kernel, the ABC likelihood \eqref{ABC_like} can be computed analytically, and the corresponding posterior distribution 
for $\theta$ is 
\begin{align}
 p_{ABC,\epsilon}(\theta|y) =& N\left(\frac{n/(1+\epsilon)}{1+n/(1+\epsilon)}\bar{y},\frac{1}{1+n/(1+\epsilon)}\right). \label{abcposterior}
\end{align}
Being able to compute the targeted posterior analytically for the VBIL approach is important.  This is because the use of a finite $\epsilon$ inflates the targeted
posterior variance compared to the truth, whereas the VB approximation can result in an error in the opposite direction (underestimation of variance
will occur in this example if we have not perfectly tuned the variance of log likelihood estimates to be constant across the parameter space).  
So apparent good performance of VBIL can sometimes result simply from a fortuitous cancellation of these errors in different directions, so it is important to
understand what distribution is being targeted by the VBIL algorithm.  

We consider $d=n=4,8$ and set $S = 100$. For VBSL  we fixed $N = 50$. For VBIL, we set the ABC tolerance parameter $\epsilon$ in \eqref{ABC_normal} as $0.1282$ and $0.1139$ for $d=4,8$ respectively. These values are chosen to ensure that (\ref{abcposterior}) only overestimates the true posterior standard deviation
by $10$\%, which is a reasonable standard of accuracy. Of course, since the summary statistic is exactly Gaussian here the synthetic likelihood method is exact.  
We set the minimum value of $N$ in VBIL to be $50$, but implement the adaptive sample size approach described in Section \ref{sec:vbil} to tune $N$ to target variances of $\log \hat{p} (y|\theta)$ of 0.1 and 0.5 and denote these two methods by $\mathrm{VBIL}_{0.1}$ and $\mathrm{VBIL}_{0.5}$ respectively.  
On average, for $d=4$,  we required approximately $N = 60$ and $N = 400$ simulations to achieve $\mbox{Var}(\log \hat{p} (y|\theta)) \leq 0.5$ and $0.1$ respectively, and  an average of $N = 250$ and $N=6500$ simulations for $d= 8$.

Note that with these specifications one iteration of the VBSL algorithm takes either the same or less computational effort than  the VBIL approaches, so that faster convergence of VBSL implies less computational effort overall.  
We set the learning rate $\rho_t = \frac{1}{5+t}$ where $t$ is the iteration number;  this form satisfies the Robbins-Monro conditions with constants hand tuned for good performance in the VBIL
approaches.  The effects of adaptive learning rates are investigated further in later examples.  
We initialize our starting point for $q(\theta)$ to be $N(\mu^{(0)},\sigma^{(0)})$ where $\mu^{(0)}$ is the mean of the observed data and $\sigma^{(0)} = 1$. We fixed the number of iterations to 100. 

\begin{figure}[!t]
\centering
\includegraphics[width=1\textwidth]{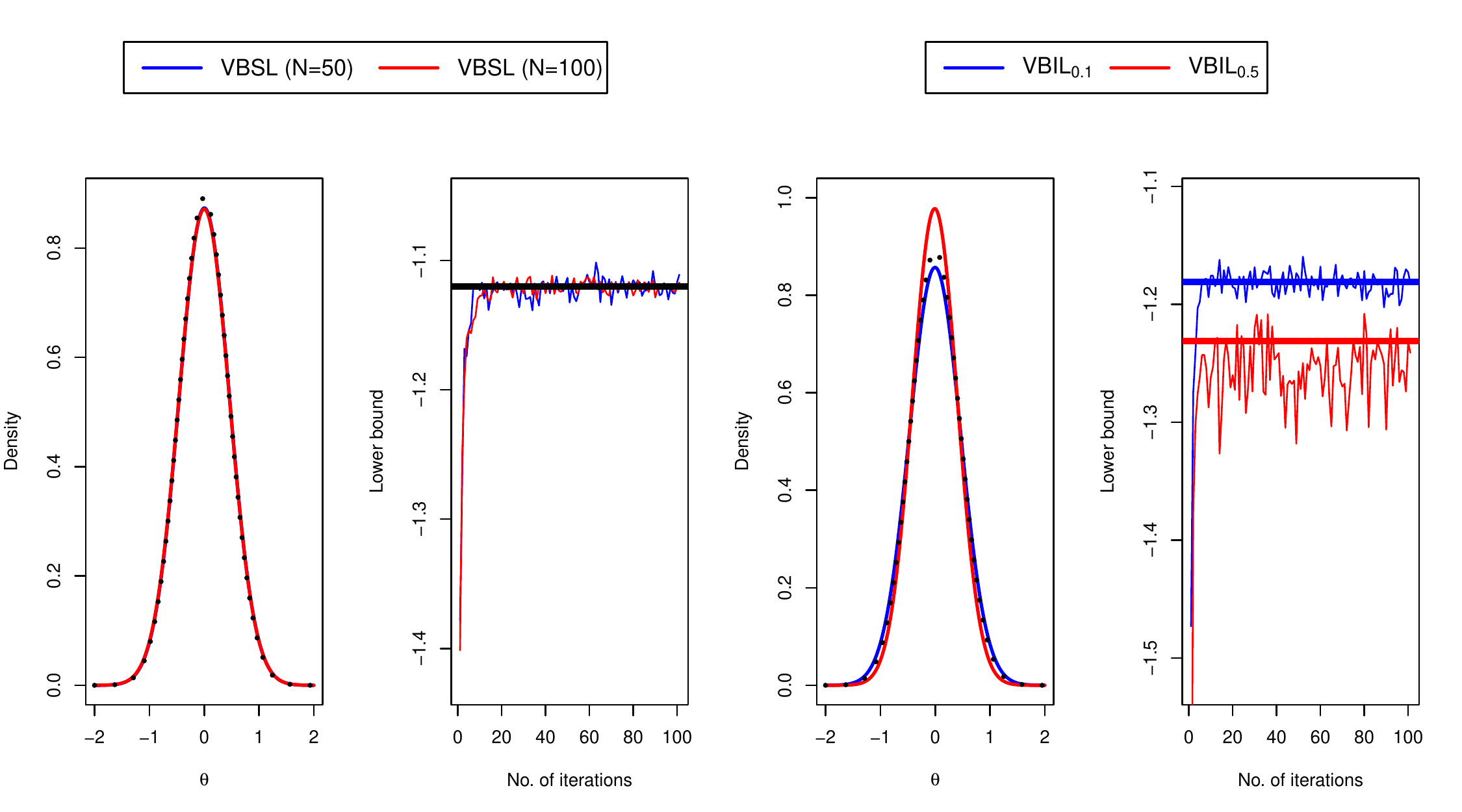}
\caption{\label{Toy2} \small Posterior distribution and variational lower bound for $d = 4$. The black dotted line on the density plot represents the true posterior distribution. The horizontal black, blue and red lines in the lower bound plot represents the analytically calculated lower bound for VBSL, $\mathrm{VBIL}_{0.1}$ and $\mathrm{VBIL}_{0.5}$ respectively. }
\end{figure}

\begin{figure}[!t]
\centering
\includegraphics[width=1\textwidth]{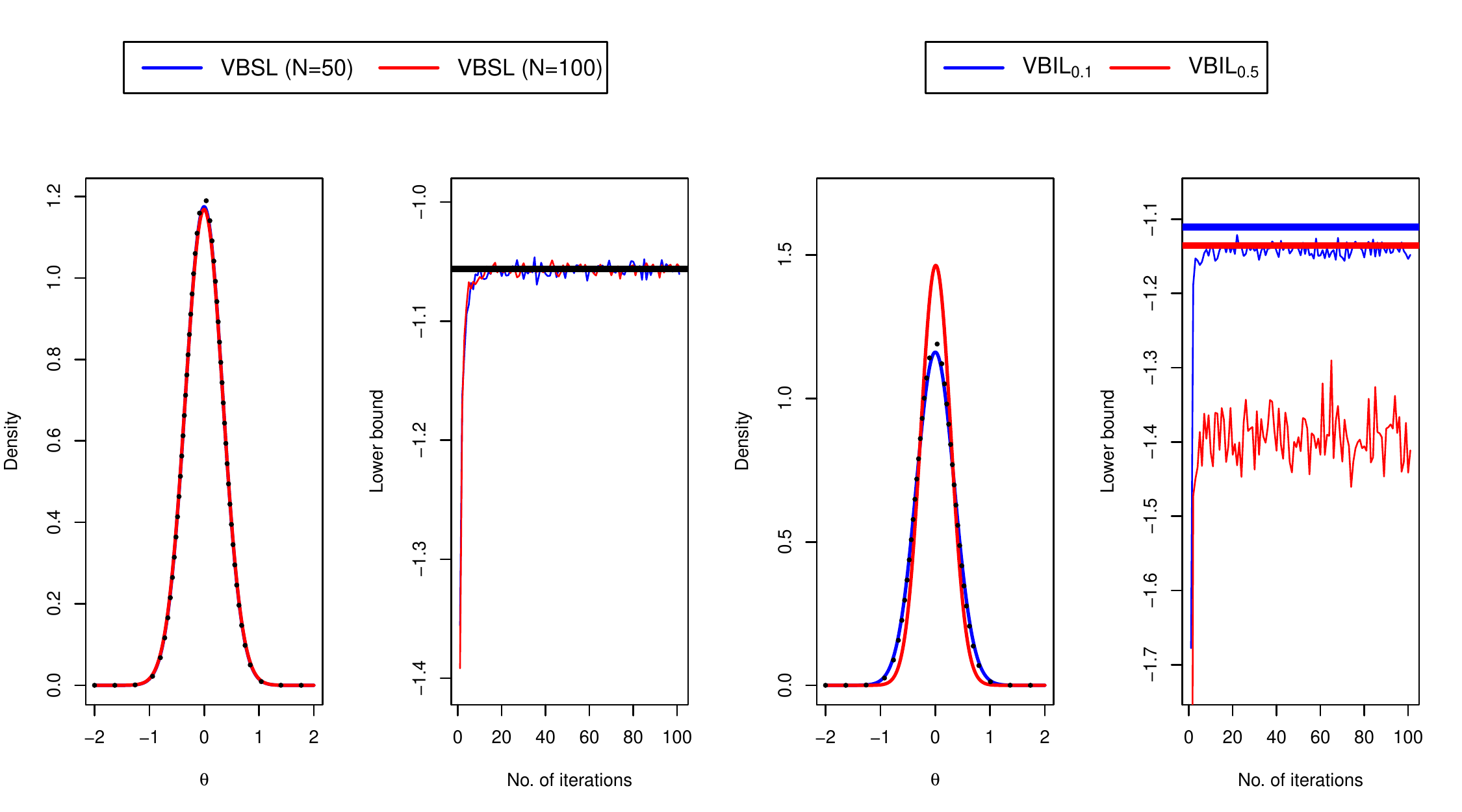}
\caption{\label{Toy3}\small Posterior distribution and variational lower bound for $d = 8$. The black dotted line on the density plot represents the true posterior distribution. The horizontal black, blue and red lines in the lower bound plot represents the analytically calculated lower bound for VBSL, $\mathrm{VBIL}_{0.1}$ and $\mathrm{VBIL}_{0.5}$ respectively. }
\end{figure}

In this toy example, it is possible to calculate the optimised variational lower bound value analytically.
In the case of  VBIL, this assumes that it is properly tuned so that the variance of the log-likelihood estimate is constant.  In particular, considering 
$1/n \log p(y) = (1/n) \log \int p(y|\theta) p(\theta) d\theta$ and replacing $ p(y|\theta)$ with the ABC or synthetic likelihood, the (scaled) lower bound is
\[
LB_{VBSL} = -\frac{1}{2} \log (2\pi) - \frac{1}{2n} \sum_{i=1}^n y_i^2 - \frac{1}{2n} \log (n+1) + \frac{1}{2n(n+1)} \left( \sum_{i=1}^n y_i \right)^2
\]
for the VBSL approach and
\begin{align*}
LB_{VBIL} &=  -\frac{1}{2} \log (2\pi) - \frac{1}{2} \log(1+\epsilon)- \frac{1}{2n (1+\epsilon)} \sum_{i=1}^n y_i^2  \\
& \qquad - \frac{1}{2n} \log \left(\frac{n}{1+\epsilon}+1\right) + \frac{(n/(1+\epsilon))^2}{2n(1+n/(1+\epsilon))}\bar{y}^2-\frac{\tau^2}{2n}
\end{align*}
for the VBIL approach, where $\tau^2$ is the (assumed constant) targeted variance for the log likelihood estimate (for the VBIL method we have used  (\ref{auglowerbound}) to derive this expression).
How close we come to attaining these analytically calculated lower bound expressions is a measure of the accuracy of the algorithm taking into
account the different likelihoods implictly being used, and also helps assess convergence of the algorithm.

Figures \ref{Toy2} and \ref{Toy3} illustrate the variational distribution of $\theta$ and the realised lower bound using VBSL and VBIL. For both $d=4$ and $8$, the VBSL approach matches the true posterior distribution (represented by the black dotted curve) and attains its analytic lower bound (represented by the black horizontal line). For  $\mathrm{VBIL}_{0.1}$ and $\mathrm{VBIL}_{0.5}$, their means match but variances differ slightly for $d=4$ and $\mathrm{VBIL}_{0.5}$ has not really converged within 100 iterations for $d=8$.  
Unsurprisingly, the performance of the VBIL approach deteriorates when the dimension of the summary statistics increases. The estimated posterior distribution deviates from the true posterior distribution, by having a smaller variance, when we set $\mbox{Var}(\log \hat{p} (y|\theta)) \leq 0.5$ for $d=8$. The performance greatly improves if we set $\mbox{Var}(\log \hat{p} (y|\theta)) \leq 0.1$. However, we observe this method requires $N = 6200$ simulations per likelihood estimate on average, which in turn would imply a much larger computational effort. In fact, we found that for $d=8$, the synthetic likelihood with $N=50$ and $\mathrm{VBIL}_{0.1}$ require 3 and 13 minutes respectively for 100 iterations. This reflects the advantage of the parametric assumptions made in the synthetic likelihood and we are able to achieve reasonable answers for less computational effort.

\subsection{$\alpha$-stable model}

We now examine the importance of adaptive learning rates within the VBSL algorithm.
$\alpha$-stable models (see, for example, \citet{adler+ft98}, Section VII) are a convenient family of heavy-tailed distributions used in a number of applications.  
Inference is challenging, since for distributions in this family there is no closed form expression for the density function.  The most common parametrization of these
distributions is in terms of a parameter $\theta=(\alpha,\beta,\gamma,\delta)^\top$, where $\alpha$ is a parameter controlling tail behaviour, 
$\beta$ controls skewness, $\gamma$ is a scale parameter and $\delta$ a location parameter. The characteristic function is
$$\phi(t)=\left\{\begin{array}{cc} \exp\left\{i\delta t-\gamma^\alpha |t|^\alpha \left(1+i\beta\tan\frac{\pi \alpha}{2}\mbox{sgn}(t)\left(|\gamma t|^{1-\alpha}-1\right)\right)\right\} & \mbox{$\alpha\neq 1$} \\
\exp\left\{i\delta t-\gamma |t|\left(1+i\beta\frac{2}{\pi}\mbox{sign}(t)\log (\gamma |t|)\right) \right\} & \mbox{$\alpha=1$} 
\end{array},\right.$$
where $\mbox{sgn}(t)$ is the sign function which is $1$ if $t>0$, $0$ if $t=0$ and $-1$ if $t<0$.  ABC methods for inference in this model
were considered by \citet{peters+sf12}, who exploit the fact that convenient simulation algorithms are available for these models.  Here we use
a univariate model, but \citet{peters+sf12} also consider the multivariate case.  
We follow \citet{tran+nk16} who apply VBIL on a dataset of size $500$ simulated from an $\alpha$-stable model with $(\alpha,\beta,\gamma,\delta)=(1.5,0.5,1,0)$.
Here we consider the performance of VBSL with the uBSL pseudo-marginal approach of \citet{price+dln16}.  

\begin{figure}[ht]
\centering
\includegraphics[width=1\textwidth]{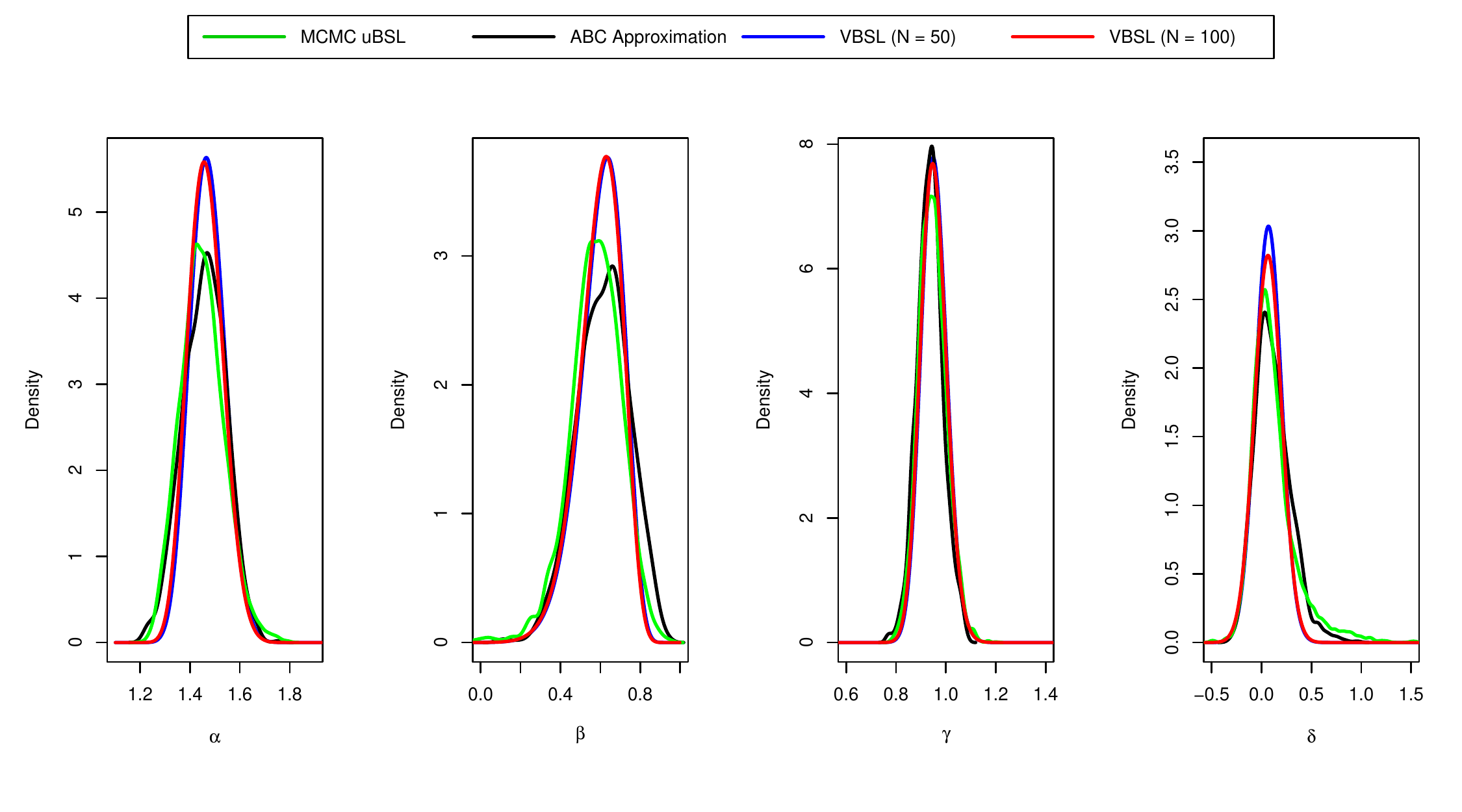}
\caption{\label{alpha_fig}\small Marginal variational posterior distributions for the four parameters $(\alpha,\beta,\gamma,\delta)$ for the $\alpha$-stable model under $N= 50$ and $N = 100$. }
\end{figure}

Similar to \citet{tran+nk16}, we enforce constraints that $\alpha\in[1.1,2]$, $\beta\in [-1,1]$ and $\gamma>0$ (e.g. \cite{peters+sf12}) through the reparametrisation
$\tilde{\theta}=(\tilde{\alpha},\tilde{\beta},\tilde{\gamma},\tilde{\delta})^\top$, with
$$\tilde{\alpha}=\log \frac{\alpha-1.1}{2-\alpha},\;\;\;\tilde{\beta}=\log \frac{\beta+1}{1-\beta}\;\;\;\tilde{\gamma}=\log \gamma\;\;\;\mbox{and }\:\:\tilde{\delta}=\delta.$$
We consider a normal prior on $\tilde{\theta}$, $N(0, I_4)$, and approximate the posterior
distribution of $\tilde{\theta}$ with a multivariate normal distribution, but report results for the posterior distribution for $\theta$ by inversion of the transformation
from $\theta$ to $\tilde{\theta}$.  
For summary statistics, we consider a point estimator of $\theta$ due to \citet{mcculloch86} and then transform this point estimator to an estimator of 
$\tilde{\theta}$.  

In implementing VBSL there are a number of algorithmic parameters to be set.  We choose $S=500$ and use 
$N=50$ and $N = 100$ to inspect the sensitivity of the VBSL towards the choice of $N$.  In this analysis we first consider the adaptive learning rate
sequence described in Section 3.  
Figure \ref{alpha_fig} shows the variational distribution of the four parameters $\alpha, \beta, \gamma$ and $\delta$. The true parameter values that are used to generate the data are recovered well -- the variational distributions are quite close to those estimated by a ``gold standard" ABC approximation with local linear adjustment, which is based on 1,000,000 generated samples, Epanechnikov kernel and a tolerance of $\epsilon=0.001$. Furthermore, we observe that the variational distribution of the parameters is quite insensitive to $N$.

\begin{figure}[ht]
\centering
\includegraphics[width=1\textwidth]{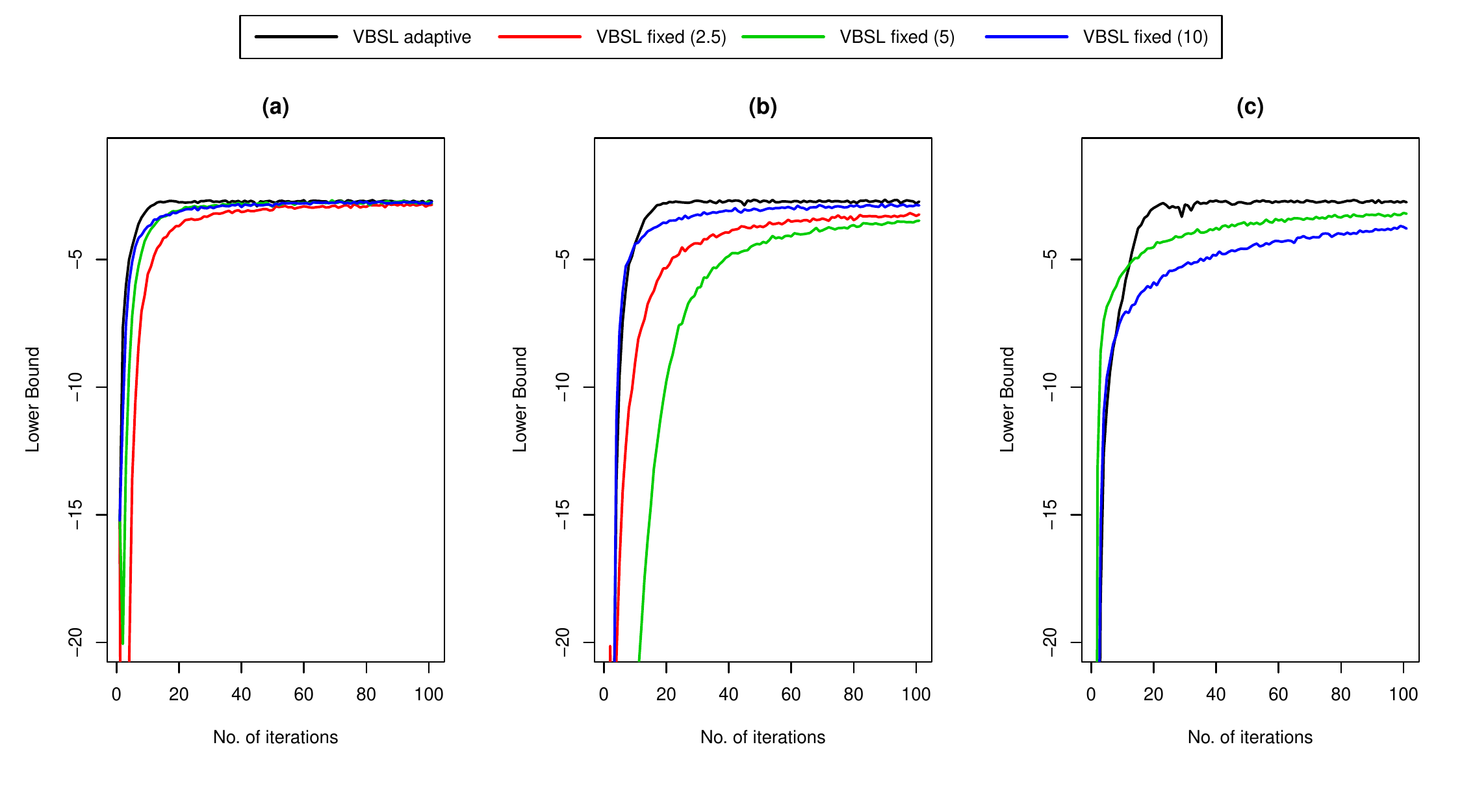}
\caption{\label{alpha_fig_two}\small  Convergence of VBSL algorithm for adaptive versus three fixed learning rate. VBSL fixed (2.5) uses $\rho_t=1/(2.5+t)$, VBSL fixed (5) uses $\rho_t=1/(5+t)$ and VBSL fixed (10) uses $\rho_t=1/(10+t)$. Convergence speed is shown for three different starting values of the variational mean for $(\alpha,\beta,\gamma,\delta)$. Namely: (a) the estimated summary statistics for the observed data, (b) $(1.5, 0.5, 3,0)$  and (c) $(1.5, 0, 2, 0)$.   }
\end{figure}

Figure \ref{alpha_fig_two} shows the convergence of the algorithm for the adaptive learning rate sequence (black line) and three fixed learning rate sequences, as a function of different starting values of the variational means for the four parameters (one ``good" starting value (a), and two poor values).  The starting variational covariance matrix is fixed at $0.04I_4$. The ``good'' variational means starting value (a) uses estimated summary statistics from the observed data. For the second and third starting values, we consider a starting variational mean of $(\alpha,\beta,\gamma,\delta)=(1.5, 0.5, 3, 0)$ and $(1.5, 0, 2, 0)$.

\begin{figure}[ht]
\centering
\includegraphics[width=1\textwidth]{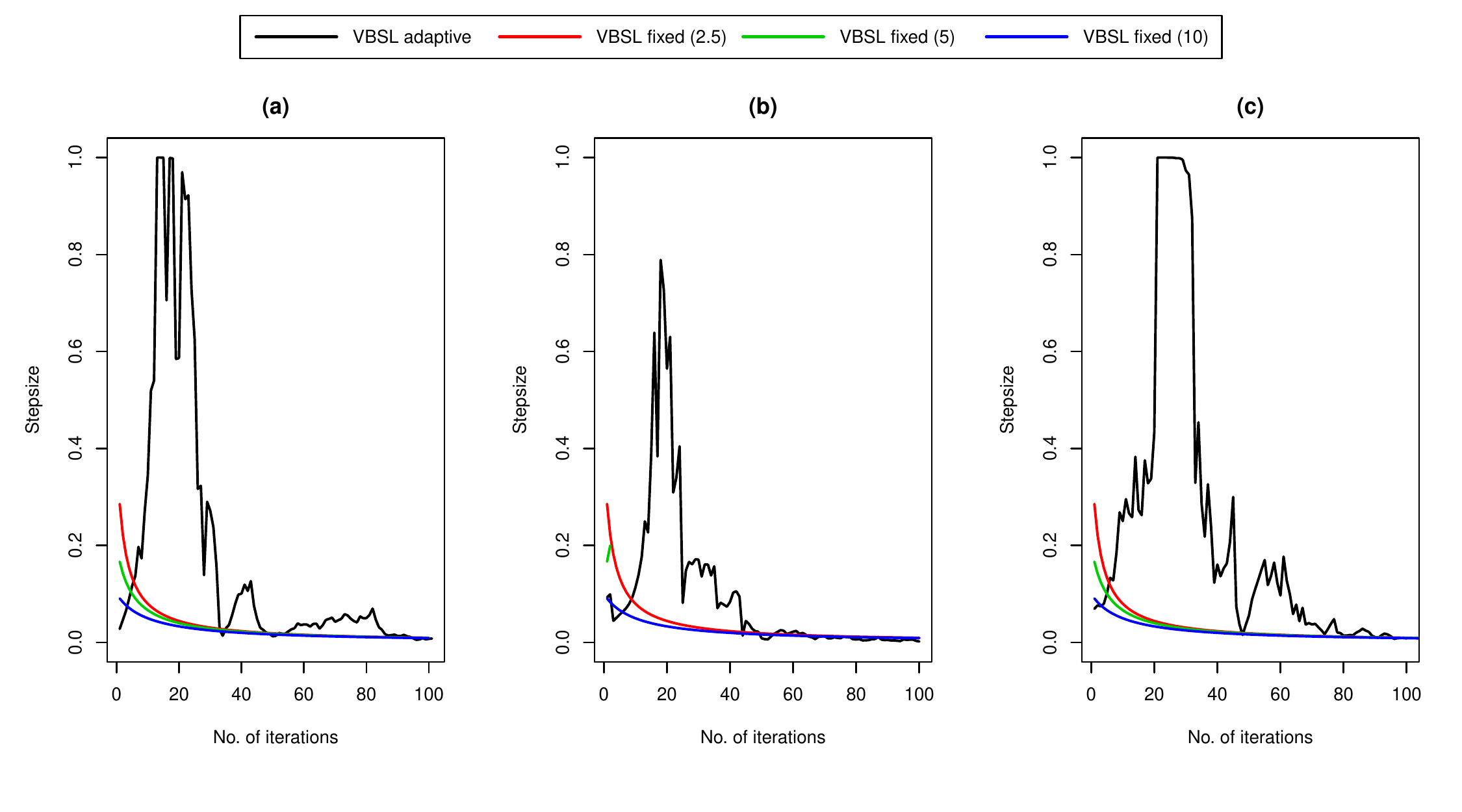}
\caption{\label{alpha_fig_three} \small Stepsize of VBSL algorithm against iteration number for adaptive versus three fixed learning rate for $\alpha$-stable model. VBSL fixed (2.5) uses $\rho_t=1/(2.5+t)$, VBSL fixed (5) uses $\rho_t=1/(5+t)$ and VBSL fixed (10) uses $\rho_t=1/(10+t)$. 
Stepsizes are shown for three different starting values of the variational mean for $(\alpha,\beta,\gamma,\delta)$. Namely: (a) the estimated summary statistics for the observed data, (b) $(1.5, 0.5, 3,0)$  and (c) $(1.5, 0, 2, 0)$. 
}
\end{figure}

Figure \ref{alpha_fig_two} demonstrates that except for the ``good" starting value (where the different learning rates perform similarly), the adaptive learning rate sequence converges much faster than all the fixed learning rates.  It is generally the case that 
the adaptive learning rate sequence is more robust to an inferior starting point. 
To support this, Figure \ref{alpha_fig_three} illustrates the step-size of the four learning rate sequences against the number of iterations. 
We observe that the adaptive learning rate frequently takes larger steps for a greater number of iterations than the fixed-rate sequences, particularly when using an inferior starting point.  

\subsection{Multivariate $g$-and-$k$ model}
\label{example:gandk}

The $g$-and-$k$ distribution \citep{rayner+m02} is another flexible family of distributions for which inference can be challenging due to the lack of a closed form density
function.  The $g$-and-$k$ distribution is defined through its quantile function, $Q(p)$, $p\in (0,1)$, where
\begin{align}
 Q(p) & = A+B\left[1+c\frac{1-\exp(-g z(p))}{1+\exp(-g z(p))}\right](1+z(p)^2)^kz(p),  \label{gandk}
\end{align}
where $z(p)=\Phi^{-1}(p)$ and $\Phi(\cdot)$ is
the standard normal distribution function.
The parameters of the family are $A$, $B>0$, $g$ and $k>-0.5$ controlling respectively the location, scale, skewness and kurtosis.  The additional parameter
$c$ is conventionally fixed at $0.8$.  Simulation from the $g$-and-$k$ model is easily done, since for $U\sim U[0,1]$, $Q(U)$ is a draw from the corresponding distribution
with quantile function $Q(p)$.  This makes ABC methods for inference attractive \citep{allingham+km09}.

Following \citet{drovandi+p11} and \citet{li+nfs15} we consider a multivariate $g$-and-$k$ model in which the copula of the distribution is a Gaussian copula.  
In particular, suppose we have independent and identically distributed multivariate observations $y_1,\dots,y_n$ where $y_i=(y_{i1},\dots,y_{iq})^\top$.  
Each $y_{ir}$ follows a univariate $g$-and-$k$ distribution marginally, $F(x;\theta_r)$ say with parameters $\theta_r=(A_r,B_r,g_r,k_r)^\top$, $r=1,\dots,q$.  
The density function and quantile function corresponding to $F(x;\theta_r)$ are written respectively as $f(x;\theta_r)$  and $Q(p;\theta_r)$.  
Dependence between components of $y_i$ is modelled using a Gaussian copula (\cite{drovandi+p11,joe97}).  
 Let $\Sigma$ be a $q\times q$ correlation matrix.  
Then the density of $y_i$ is
\begin{align}
 f(y_i;\theta) & = |\Sigma|^{-1/2}\exp\left(\eta_i^\top(I-\Sigma^{-1})\eta_i\right)\prod_{j=1}^q f(y_{ij};\theta_j) \label{gkcopula}
\end{align}
where $\eta_i=(\eta_{i1},\cdots,\eta_{iq})^\top$ with $\eta_{ir}=\Phi^{-1}(F(y_{ir};\theta_r))$.  This density cannot be computed in closed form because
the $g$-and-$k$ marginals are not available in closed form.  However, it is easy to simulate from the model.  
Simulation from the Gaussian copula based model (\ref{gkcopula}) is easily achieved by generating $Z\sim N(0,\Sigma)$ and transforming $Z$ to $(Q(\Phi(Z_1);\theta_1),\dots,Q(\Phi(Z_q);\theta_q))^\top$. For summary statistics, we follow \citet{drovandi+p11} and use 
\[
S_{A_r} = E^{(r)}_4, S_{B_r} = E^{(r)}_6 - E^{(r)}_2, S_{g_r} = \frac{E^{(r)}_7 - E^{(r)}_5 + E^{(r)}_3 - E^{(r)}_1}{S_{B_r}}, S_{k_r} = \frac{E^{(r)}_6 + E^{(r)}_2 - 2 E^{(r)}_4}{S_{B_r}},
\]
where $E^{(r)}_j$ is the $j$-th octile of the data $(y_{1r},...,y_{nr})$, for the model parameters and the robust normal scores correlation coefficient \citep{fisher+y48} for each of the correlation parameter in the off-diagonal entries of the copula correlation matrix $\Sigma$. 

The model (\ref{gkcopula}) has marginal parameters $\theta_1,\dots,\theta_q$, as well as the copula correlation matrix $\Sigma$.  It will be convenient
to work with an unconstrained parametrisation of $\Sigma$.  We will use a spherical parametrisation 
(see \citet{pinheiro+b96}, Section 2.3) and only consider the cases $q=2, 3$.  
For $q=2$, we let $w^{(2)}$ be an unconstrained real parameter and $\gamma^{(2)}=\pi/(1+\exp(-w^{(2)}))$.  We parametrise $\Sigma$ in terms of $\gamma^{(2)}$ by considering the 
Cholesky factorisation of $\Sigma$, $\Sigma=LL^\top$, and letting
$$L=\left[\begin{array}{cc} 1 & 0 \\ \cos(\gamma^{(2)}) & \sin(\gamma^{(2)}) \end{array} \right].$$
For $q=3$, we let $w^{(3)}=(w^{(3)}_1,w^{(3)}_2,w^{(3)}_3)^\top$ where the elements of $w^{(3)}$ are unconstrained
real parameters, define $\gamma^{(3)}_j=\pi/(1+\exp(-w^{(3)}_j))$, $j=1,2,3$ and parametrise the Cholesky factor $L$ of $\Sigma$ as
$$L=\left[\begin{array}{ccc} 1 & 0 & 0 \\ \cos(\gamma^{(3)}_1) & \sin(\gamma^{(3)}_1) \\
\cos(\gamma^{(3)}_2) & \sin(\gamma^{(3)}_2)\cos(\gamma^{(3)}_3) &
\sin(\gamma^{(3)}_2)\sin(\gamma^{(3)}_3) \end{array}\right].$$
For both $q=2$ and $q=3$ the entries of $w^{(q)}$ are given independent normal priors, $N(0,1.75^2)$.  
For the marginal parameters $\theta_r$ we adopt independent priors for different components $r$.  Reparametrising as $\tilde{\theta}_r=(\tilde{A}_r,\tilde{B}_r,\tilde{g}_r,\tilde{k}_r)^\top$
where
$$\tilde{A}_r=10 \log\frac{A_r+0.1}{0.1-A_r}\;\;\;\tilde{B}_i=\log\frac{B_r}{0.05-B_r}\;\;\;\tilde{g}_r=\log \frac{g_r+1}{1-g_r}\;\;\;\tilde{k}_r=\log \frac{k_r+0.2}{0.5-k_r},$$
we adopt a normal prior, $N(0,4I_4)$ for $\tilde{\theta}_r$.  

We fit models with $q=1,2,3$ dimensions, with corresponding dimensions of the parameter space being $4$, $9$ and $15$ respectively, to investigate how two different
implementations of  
VBSL perform as the dimension increases.  We parametrise the variational distribution in terms of the Cholesky factor of
the precision matrix and compare the natural gradient implementation and an adaptive step size, with the approach based on the ordinary gradient and
per parameter adaptive step sizes chosen according to the ADADELTA method of \citet{zeiler12}.  
The data we use consists of foreign currency exchange log daily returns against the Australian dollar (AUD) 
for 1,757 trading days between June 1, 2007 and 31 December, 2013 \citep{rba14}.  We consider data for 3 foreign currencies, the US dollar (USD), Japanese Yen (JY)
and the Euro (EUR).  Our univariate model uses just the USD, the $q=2$ model uses the USD and JY, and the $q=3$ model uses all 3 currencies.  

\begin{figure}[!ht]
\centering
\includegraphics[width=1\textwidth]{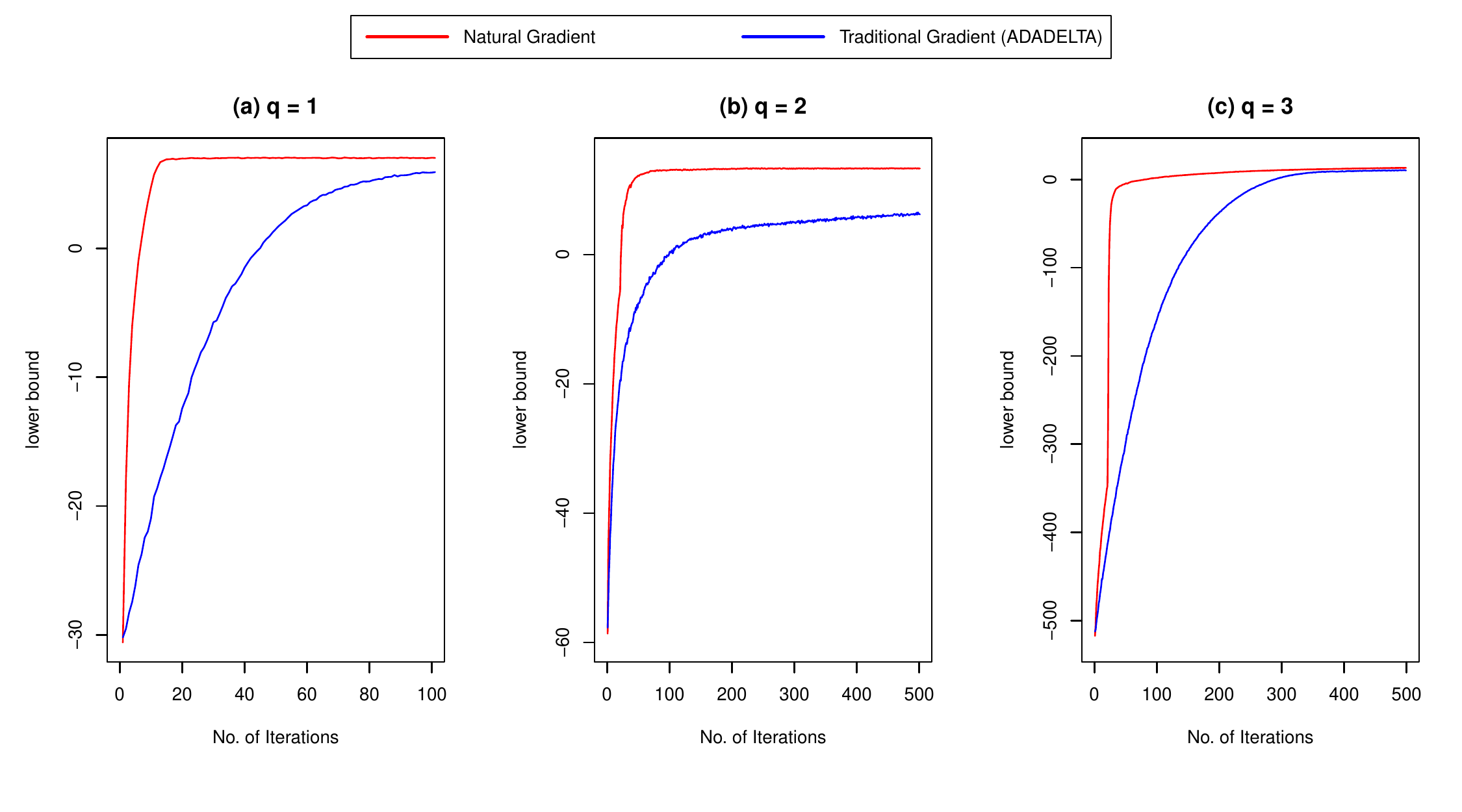}
\caption{\label{fig:q1q2_grad}\small  Lower bound against the number of iterations using the ADADELTA traditional gradient approach and our proposed adaptive natural gradient approach for the multivariate $g$-and-$k$ model. We set $N = 100$ and $S = 500$ for $q = 1, 2$, and $N = 500$ and $S = 500$ for $q = 3$.   }
\end{figure}

We set the starting values for the variational means of $(\tilde{A}_r,\tilde{B}_r,\tilde{g}_r,\tilde{k}_r)$ as $(0, -1.5, -0.5,0)$ and the corresponding variational variances as $(0.0001, 0.001, 0.1,0.1)$ for $r = 1, 2$. In the  $q=3$ dimensional model (a $15$ dimensional parameter), the starting value is based on the variational optimisation for the $q = 2$ model. In particular, we use the final variational mean and covariance matrix from $q = 2$ and set the starting value for the variational mean of $(\tilde{A}_3,\tilde{B}_3,\tilde{g}_3,\tilde{k}_3)$ as $(0, -1.5, -0.5,0)$ and the corresponding variational posterior variances as $(0.0002, 0.001, 0.1,0.1)$. For the other algorithmic parameters, we set $N = 100$ and $S = 500$ for $q = 1, 2$ and $N = 500$ and $S = 500$ for the highest dimensional example, $q = 3$. A larger $N$ seems to be required when dealing with higher dimensional summary statistics, particularly in the initial stages, when trying to estimate likelihoods for many parameter values out in the tails of the likelihood can result in highly variable estimates. The natural gradient approach is more sensitive to this effect than the ordinary gradient approach, although the natural gradient converges faster if 
a large enough $N$ is used. 

Figure \ref{fig:q1q2_grad} shows the progress of the lower bounds for the two different schemes. We found that the adaptive natural gradient approach converges quite rapidly for all models, while the ordinary gradient requires a much larger number of iterations.

\subsection{Cell motility example}

\citet{price+dln16} consider an analysis involving a stochastic model of collective cell spreading.  The model contains two parameters: $P_m \in (0,1)$ (the probability that a cell moves to a neighbouring location in a small time step) and $P_p \in (0,1)$ (the probability that a cell gives birth to a daughter that is placed in a neighbouring location in a small time step).  \citet{price+dln16} consider a simulated dataset involving a time series of binary matrices where a 1 denotes the presence of a cell at a particular location.  This dataset is condensed into a 145 dimensional summary statistic, which is difficult to accommodate in conventional ABC settings. They obtain significant computational advancements using 
a pseudo-marginal synthetic likelihood approach --  
however, the posterior inference remains time consuming.  For more details about this application see \citet{price+dln16} and the references therein.

\begin{figure}[!ht]
\centering
\includegraphics[width=1\textwidth]{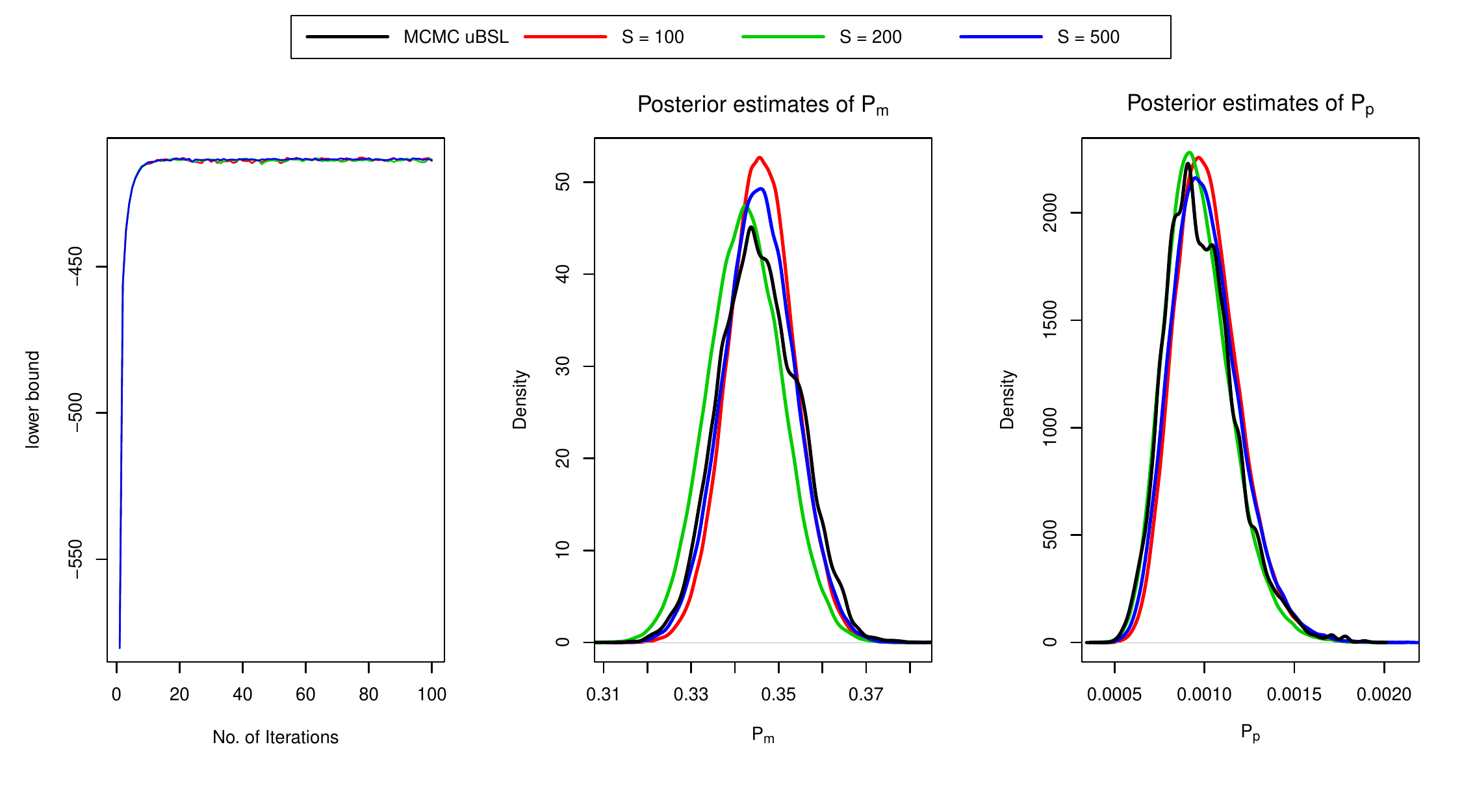}
\caption{\label{fig:cell_lb}  \small Convergence of the VBSL algorithm for the cell motility analysis with $N=1000$ and different values of $S$.  }
\end{figure}

For the variational distribution we use a bivariate normal distribution on the logit of the parameter space.  We run VBSL with $N=1000$ using our
adaptive natural gradient algorithm and $S=100, 200, 500$.  We note that \citet{price+dln16} find that the best choice of $N$ in terms
of computational efficiency in the context of their pseudo-marginal algorithm is $N=5000$ (out of the trialled values of 2500, 3750, 5000, 7500 and 10000).
Figure \ref{fig:cell_lb} shows plots of the variational lower bound against algorithm iteration
 and the posterior density of $P_m$ and $P_p$.  
We observe that with the adaptive scheme the VBSL methods converge rapidly and their posterior estimates are similar to the pseudo-marginal synthetic likelihood approach. However, the total computational effort involved is much reduced compared to the MCMC application considered in \citet{price+dln16}.  
In the MCMC scheme, 50,000 iterations with $N=5000$ requires $250$ million simulations of the summary statistics.  On the other hand, with $S=100$, and given that our VBSL
scheme converges in about $20$ iterations (and taking into account a further $5$ iterations used in initialization of the adaptive step size) 
the number of summary statistic simulations required is about $2.5$ million for VBSL, so that the computational requirement is about 100 times less.  

\section{Discussion}

We have introduced a new VB approach to likelihood free inference based on unbiased estimation of the log likelihood in the situation
where the summary statistic is approximately Gaussian.  In situations where the approximate Gaussian assumption holds, the methods are able to 
achieve good accuracy with much less computational effort than conventional ABC or synthetic likelihood methods.  
A focus of our future work will be making the form of the variational posterior more flexible (i.e. non-Gaussian) and implementing suitable variance reduction
methods in estimating stochastic gradients in this situation.  The local expectation gradients (LEG) framework of \citet{titsias+l15} may be particularly
useful here.  

\section*{Appendix A}

This appendix explains the parametrization of the variational distribution and computation of the information matrix $I_F(\lambda)$ in Algorithm 1. Most of the notations, i.e. $\mbox{vec}, \mbox{vech}, D_d^+$ and $D_d$, can be found in Section \ref{Repara_update}. We also write $\mbox{vec}^{-1}(a)$ for the inverse operation that takes a vector $a$ of length $d^2$ and makes a $d\times d$ matrix by filling
up the columns from left to right from the elements of the vector.

Suppose that $q_\lambda(\theta)$ represents our multivariate normal variational posterior approximation.  $\lambda$ will denote the natural
parameters in the exponential family representation of the density, given below.  Writing $\mu$ and $\Sigma$ for the mean and covariance matrix of
$q_\lambda(\theta)$, we have \citep{wand14}
$$\lambda=\left[\begin{array}{l} \lambda_1 \\ \lambda_2 \end{array} \right]=\left[ \begin{array}{c} \Sigma^{-1}\mu \\ -\frac{1}{2}D_d^\top\mbox{vec}(\Sigma^{-1}) \end{array} \right]$$
and we can write $\mu$ and $\Sigma$ in terms of $\lambda$ as 
$$\mu=\mu(\lambda)=-\frac{1}{2}\left\{\mbox{vec}^{-1}({D_d^+}^\top\lambda_2)\right\}^{-1}\lambda_1,\;\;\;\;\; \Sigma=\Sigma(\lambda)=-\frac{1}{2}\left\{\mbox{vec}^{-1}({D_d^+}^\top\lambda_2)\right\}^{-1}.$$
The exponential family representation is  
$$q_\lambda(\theta)=\exp\left(T(\theta)^\top \lambda-Z(\lambda)\right),$$
where $T(\theta)$ is the sufficient statistic
$$T(\theta)=\left[ \begin{array}{c} \theta \\ \mbox{vech}(\theta \theta^\top) \end{array} \right]$$
and $Z(\lambda)$ is the appropriate normalizing constant.  
\citet{wand14} shows that with $I_F(\lambda)$ defined as $\mbox{Cov}_\lambda(T(\theta))$, where $\mbox{Cov}_\lambda$ denotes the covariance computed using
expectation with respect to $q_\lambda(\theta)$, then (again using similar notation to \citet{wand14})
$$I_F(\lambda)^{-1}=\left[ \begin{array}{cc} \Sigma^{-1}+M^\top S^{-1}M & -M^\top S^{-1} \\ -S^{-1}M & S^{-1} \end{array} \right],$$
where $M=2D_d^+ (\mu\otimes I_d)$ and $S=2 D_d^+ (\Sigma\otimes \Sigma) {D_d^+}^\top$ and $\otimes$ denotes the Kronecker product.  
Finally
$$\nabla_\lambda \log q_\lambda(\theta)=\left[ \begin{array}{c} \theta-\mu \\ \mbox{vech}(\theta\theta^\top-\Sigma-\mu\mu^\top ) \end{array} \right].$$

\section*{Acknowledgements}

Victor Ong and David Nott were supported by a Singapore Ministry of Education Academic Research
Fund Tier 2 grant (R-155-000-143-112). Christopher Drovandi was supported by an Australian Research
Council's Discovery Early Career Researcher Award funding scheme (DE160100741). SAS was supported by the Australian Research Council (DP160102544).

\bibliographystyle{apalike}
\bibliography{references}

\end{document}